\documentclass[%
 reprint,
 amsmath,amssymb,
 aps,
]{revtex4-2}

\usepackage{graphicx}
\usepackage{dcolumn}
\usepackage{bm}
\usepackage{lipsum}
\usepackage[colorlinks=true]{hyperref}

\usepackage{graphicx}
\graphicspath{ {images/} }
\usepackage{dcolumn}
\usepackage{bm}
\usepackage{mathtools}
\usepackage{xcolor}

\usepackage{autobreak}

\newcommand{\dd}{{\mathrm d}}

\usepackage{xcolor}

\newcommand{\tro}{\mathrm{tr} \, \mathcal{O}}

\begin{document}

\title{Measuring qualitative change: A variational score for tracking dynamical shifts in partial differential equations}
\author{Joseph J. Pollacco}
\thanks{Authors contributed equally}
\author{Jonathan Wong}
\thanks{Authors contributed equally}
\author{Navonil Neogi}
\author{Callum Simpson}
\author{Eloisa Bentivegna}
\email{eloisa.bentivegna@ibm.com}
\affiliation{%
 IBM Research\\
 Keckwick Lane, Daresbury WA4 4AD, United Kingdom
}%

\begin{abstract}
Partial differential equations (PDEs) regulate the behaviour
of countless spatiotemporal systems in the physical and life
sciences. In many cases, they encode the coupling
between the system's degrees of freedom, leading to nonlinear
equations whose solution space is challenging to 
explore exhaustively. Systematic approaches to PDE model exploration
are a holy grail of computational science.
In this article, we formulate a criterion for increasing the
diversity of a search campaign, based on the PDE residual behaviour
under solution deformation. We develop a practical formalism to compute
this property and illustrate its role in a few
cases of interest.
\end{abstract}

\pacs{Valid PACS appear here}
\maketitle


\section{Introduction}

Nonlinear dynamical systems are known to exhibit a vast range of 
behaviours, depending on the system's phase-space structure~\cite{strogatz2018nonlinear,bountis2012complex}. 
Discovering such behaviours exhaustively and efficiently at scale
is a grand challenge in many areas of computational science~\cite{Feynman1964VolumeII}.
For instance, the limited diversity of drug candidates obtained via 
{\it in-silico} discovery is a barrier to safer and more effective 
clinical trials~\cite{Sadybekov23}; similarly, a comprehensive 
understanding of how high-dimensional, tightly integrated systems
behave could help explain brain function and disease~\cite{bassett17};
the computational discovery and characterisation of "novel climates"~\cite{doi.org/10.1890/070037}
will help understand ecosystem evolution on Earth; the generation
of detailed cosmological model ensembles, and, in particular, the availability of
efficient strategies for parameter exploration will be crucial
for making sense of the burgeoning volume of space data
without systematic biases~\cite{2022arXiv220307347A}.

Whilst the challenge of solution-space exploration is transversal to
many computational disciplines, model-agnostic measures of changes in dynamical behavior
(such as changes in the characteristic spatio-temporal
variability of a system~\cite{doi:10.4169/000298910X496723}, pattern formation~\cite{2003SJADS...2..487L}, or the emergence
of singularities ~\cite{PhysRevLett.130.244002, PhysRevD.103.044016})
are difficult to define.
The common, practical approach to solution discovery involves 
identifying specific properties that are relevant for the application 
of interest, and searching the solution space for critical points
in these quantities. When solution stability is important, for instance,
the algorithms of~\cite{tailleur2007probing, 2011JSP...145..787G} can be used, where
Lyapunov exponents are used as an objective driving a solution 
discovery based on population-based optimisation. Searches for
quantitative extremes in predefined observables can similarly be 
instrumented~\cite{2022arXiv220402488P}.
The use of statistical frameworks to quantify extremes in 
observables over solution ensembles is also well established
\cite{lucarini2016extremes}.

In the following, we focus on models based on partial differential 
equations (PDEs), and endeavour to find markers of dynamical transitions 
that solely depend on the form of the governing PDE, with no 
assumption on the character of the transition or the observables
that may be affected. Inspired by a variational condition on the accuracy
of a certain class of surrogate PDE models~\cite{DBLP:conf/iclr/Bentivegna23},
we are able to formulate a {\it deterministic} approach
for driving solution-space explorations efficiently towards new
dynamical regimes. The key quantity driving the exploration is a
variational score, defined via functional derivatives of the PDE residual.
We include a discussion of two other measures of solution variability
and their complementarity to the score proposed here.

In Section~\ref{sec:var}, we define the model equations and its solution
space, and introduce the variational score. We test this approach 
on five differential systems in~\ref{sec:examples}. The ability 
to read solution-manifold properties from the variational structure of
a PDE is intriguing, and we summarise the power of this approach and a
number of promising open directions in~\ref{sec:concl}.

\section{A score for dynamical transitions}
\label{sec:var}

\subsection{PDEs and their solution manifolds}
\label{sec:sman}
We consider a system of partial differential equations, defined over a spatial
domain $\Omega$ and temporal interval $[t_i,t_f]$, and supplemented with 
any required initial and boundary conditions:
\begin{eqnarray}
&&r^e(\partial^p_{\{c_p\}}u^a,x^b;\mu^d)=0 \qquad p=0,\dots, p_{\rm max}, \label{eq:pde} \\
&&u^a|_{t_i}=v^a \quad \qquad {v^a}:\Omega \times \mathcal{M} \to \mathbb{R}^n, \label{eq:IC}\\
&&u^a|_{\partial \Omega} = {w^a} \qquad {w^a}:\partial\Omega \times [t_i,t_f] \times \mathcal{M} \to \mathbb{R}^n, \label{eq:BC}
\end{eqnarray}
where $r^e$ are the individual residuals of the equations in the system.
The set $\{c_p\}$ represents any combination of
$p$ spacetime indices.
We will concern ourselves with the case where some aspects of the problem
(such as the equation itself, or any of the conditions (\ref{eq:IC})-(\ref{eq:BC}))
are formulated parametrically, with the parameters $\mu^d \in \mathcal{M}$.
Unless otherwise noted, we will assume that both $u^a$ and all relevant
derivatives are defined and finite throughout the domain.

Under a fairly general set of assumptions, $u^a$ lives in a 
Sobolev space $W^{l,p}(\Omega)$. The set of solutions of equation
(\ref{eq:pde}) forms a submanifold $U \subseteq W^{l,p}(\Omega)$, 
spanned by the coordinates ${\mu^d}$ (these may, in general, 
be made up of multiple coordinate patches, depending on $U$). 
Finding the parametric representation for solutions of a PDE
that optimally simplifies the search of $U$ is generally a complex, open-ended problem.

\subsection{The residual functional and its variations}
The main question explored by this work is how one can explore $U$ 
efficiently, detecting when a shift occurs
in the dynamical properties of a group of solutions and another
sector is entered. In other words, we aim to formulate a procedure
for iteratively learning the equivalent of a linear system's Poincar{\'e} diagram.

Ideally, the procedure would be formulated as the optimisation of a single,
scalar-valued score, which:
\begin{enumerate}
    \item exhibits outlier values for transition solutions,
    \item is easily searchable for such outlier values, and
    \item does not require the transition to entail a quantitative
    shift in the solution's norm, i.e.~can detect qualitative
    changes.
\end{enumerate}

As illustrated below, we observe that such a score can be defined based on
the sensitivity of the PDE residual to arbitrary deformations of its 
solutions. Solutions at the boundaries of known dynamical shifts are
shown to correspond to sharp changes in this score, and the landscape
around these boundaries is sufficiently smooth that its location could
be identified even from a distance.
The use of a variational approach is inspired by the results of \cite{DBLP:conf/iclr/Bentivegna23}, 
where the second variation of a PDE residual, calculated on a dataset of its 
solutions, is shown to impact the observational bias of an encoder-decoder 
system trained thereon, leading to models with smaller reconstruction error
near training data with a low variational score.
Whether an information-theoretical link exists between this score and the information
which an individual PDE solution carries about its $U$ neighbourhood 
is a compelling question, 
which we will address in future work; in this article, we demonstrate the utility of examining the  score's landscape in five different 
applications.

In order to estimate a PDE's residual violation around one of its
solutions and for any arbitrary deformation, we define a 
residual functional and expand it in the variational sense, focusing
on its leading-order term $\delta^2 S$ and its generator $\mathcal{O}$.

Given a PDE (\ref{eq:pde}), we define the residual functional $ R^e[{u^a}]$ as 
the $L_2$-norm over spacetime of the PDE residual $r^e$:
\begin{eqnarray}
    R^e[u^a] &=& \left [ \int_{\Omega \times [t_i,t_f]} |r^e|^2 {\rm d} x^b  \right ]^{1/2}. \label{eq:resfunc}
\end{eqnarray}
We will assume that this norm is finite, as well as all similar expressions below.
As $r^e$ is identically zero for solutions of (\ref{eq:pde}),
so is $R^e$. Although we could in principle model $R^e$ in a neighborhood of the
solution manifold through an expansion in the functional derivatives
of $R^e$, notice that, because of the square root in 
(\ref{eq:resfunc}), $R^e$ may not be differentiable with respect to $u^a$.
This problem is easily circumvented through a change of variable 
$S^e = (R^e)^2$. We can then use the expansion~\cite{1994PhRvA..50.4593E}:
\begin{equation}
    S^e[u^a+\epsilon \eta^a] = S^e[u^a]+\epsilon \delta S^e[u^a, \eta^a]
    +\frac{\epsilon^2}{2} \delta^2 S^e[u^a, \eta^a] + O(\epsilon^3).
\end{equation}
As $S^e$ is both zero on the solution manifold and non-negative elsewhere,
the solution manifold is a minimum for this functional and its
first functional derivative $\delta S^e$ is also zero there. The leading-order contribution
to $S^e$ around the solution manifold is then the second variation
$\delta^2 S^e$. 
This term measures how much equation violation results from a 
given solution deformation ${\eta^a}$. Based on the same argument,
$\delta^2 S^e$ is non-negative, and can be expanded to:
\begin{equation}
    \delta^2 S^e = \int \frac{\partial^2 (r^e)^2}{\partial(\partial^p_{\{c_p\}}u^a)\partial(\partial^s_{\{g_s\}}u^f)} \partial^p_{\{c_p\}}\eta^a\partial^s_{\{g_s\}}\eta^f {\rm d} x^b,
\end{equation}
where a sum over $a$, $p$, $f$, and $s$ is implied.
This can be rewritten as \cite{kot2014first}:
\begin{equation}
\label{om}
    \delta^2 S^e = \int \omega^e(\partial^p_{\{c_p\}}\eta^a) {\rm d} x^b,
\end{equation}
where the $\omega^e$ are homogeneous forms of 
degree two in the deformation $\eta^a$ and its derivatives.
They can be rewritten as \cite{kot2014first}:
\begin{equation}
\label{eq:om2}
    2 \omega^e =
    \frac{\partial{\omega^e}}{\partial(\partial^p_{\{c_p\}}\eta^a)}\partial^p_{\{c_p\}}\eta^a.
\end{equation}
This expression quantifies, to lowest order, the sensitivity of the
residuals $r^e$, calculated at $u^a$, to deformations
$u^a \to u^a + \epsilon \eta^a$.

The quantity $\delta^2S^e$ can then be expressed as the integral of a 
sum of bilinear operators ${\cal O}^e$, applied twice to the 
deformations $\eta^a$. This
can be made more explicit by leveraging, once again, integration
by parts:
\begin{equation}
    \delta^2 S^e = \int \eta^{a \top} \mathcal{O}^e_{af} \eta^f \; {\rm d} x^b,
\end{equation}
with
\begin{equation}
\label{eq:O-general-expression}
\mathcal{O}^e_{af} = \sum_{s \geq p} \frac{(-1)^p}{1+\delta_{sp}} \partial_{\{c_p\}} \left (\frac{\partial^2 \omega^e}{\partial(\partial^p_{\{c_p\}}\eta^a)\partial( \partial^s_{\{g_s\}}\eta^f)} \partial^s_{\{g_s\}}(\cdot) \right ),
\end{equation}
where is $\delta_{sp}$ is the Kronecker delta between $s$ and $p$.
An $\eta$-independent estimate of the magnitude of $\delta^2S^e$ over the equation index $e$
can then be obtained by solving the coupled eigenproblem for all
the $\mathcal{O}^e$, i.e.~solving the eigenproblem for the operator
$\mathcal{O}$, defined as:
\begin{equation}
    \mathcal{O} = \sum_e \mathcal{O}^e,
\end{equation}
acting on the cumulative deformation, defined as the column vector
of all deformations:
\begin{align}
    \eta &= \begin{pmatrix}
           \eta^a
         \end{pmatrix}.
\end{align}

Locally, $\delta^2 S = \sum_e \delta^2 S^e$ represents the local curvature of the solution manifold
in the direction of $\eta$;  $\mathcal{O}$'s eigenvalues quantify the curvature
of this space along its principal directions, and its trace represents the scalar curvature around $u^a$. We propose to use
$\tro$ as a measure of solution diversity in the neighbourhood of a given
$u^a$, as a shallower $\delta^2 S$ means that larger deformations are permitted
for the same level of violation of $S$.

An interesting parallel exists with the calculation of transition probability 
between metastable states in stochastic systems. There, the probability
of a given transition path to be realised is expressed as an action functional, 
and its second variation measures the volume of the transition tube, i.e.~its
cumulative probability~\cite{freidlin1983random, Schorlepp2023ScalableMF}. This quantity is estimated using a diagonalisation 
procedure, just as above. Whilst the mathematical structure of these two problems
is closely related, their physical interpretation is quite distinct. We reiterate that
we aim to use $\delta^2 S$ as a measure of solution diversity in a neighborhood
of the solution manifold. The PDE problems we consider are deterministic and noise-free.

We conclude this section by observing two interesting properties of
$\delta^2 S^e$:
\begin{itemize}
    \item For a linear problem, 
        \begin{eqnarray}
            S^e[u^a+\epsilon \eta^a] = S^e[u^a] + \epsilon S^e[\eta^a] = \epsilon S^e[\eta^a],
    \end{eqnarray}
    where the last equality holds if $u^a$ is a solution. The
    residual violation due to the deformation $\eta^a$ thus
    depends only on $\eta^a$ -- in other words, linear PDEs 
    have a constant $\delta^2 S^e$
    throughout the solution space, so that equal deformations
    correspond to equal increases in $S^e$. If $\eta^a$ is induced
    by a change in $\mu^d$, this implies that solution
    sensitivity to the controls is constant throughout the solution
    manifold, exactly as expected.

    \item It is easy to prove that $\delta^2 S^e$ can be expressed in
    terms of the square of the integrand of $\delta S^e$ when $u^a$ is a 
    solution. This would simplify the construction of expressions (\ref{om}) and (\ref{eq:om2});
    the complexity would be shifted to the integration by parts 
    required to obtain (\ref{eq:O-general-expression}). Depending on which specific quantity is of
    interest, this property can be practically useful.

\end{itemize}

In the next section, we will apply this formalism to five
differential problems of scientific interest across a range of application domains.

\section{Examples}
\label{sec:examples}

\subsection{Particle in an effective potential}
To illustrate the concept of solution transition and observe
the behaviour of $\cal{O}$ around the corresponding regions in 
the solution manifold, we begin with a simple example.

Let us consider the behaviour of a particle 
in an effective potential, given by the well-known equation:
\begin{eqnarray}
    \dot r^2(t) &=& E - V(r(t),L). \label{eq:rdot}
\end{eqnarray}
where, here and in the following, a dot denotes differentiation
with respect to time, and $L$ is a free parameter.

Defining $s = \dot r^2(t) - E - V(r(t),L)$, we can easily calculate the second variation of the residual of
(\ref{eq:rdot}):
\begin{eqnarray}
    \delta^2 S &=& \int \left (\frac{\partial^2 s}{\partial r^2} \eta^2
    + 2 \frac{\partial^2 s}{\partial \dot r \partial r} \eta \dot \eta
    +\frac{\partial^2 s}{\partial \dot r^2} \dot\eta^2 \right ) \dd t\\
    &=& \int (2 V'^2 \eta^2 + 4 V' \dot r \eta \dot \eta + 8 \dot r^2 \dot \eta^2) \dd t,
\end{eqnarray}
where a prime denotes derivative with respect to $r$, and we 
have used equation (\ref{eq:rdot}) to simplify some of 
the terms in the last line. If we consider trajectories for 
$t \in [0,T]$, and discretize this interval into $N$ points
$t_i$, it is straightforward to prove that
\begin{equation}
    \tro = 2 \sum_{i=1}^{N} (V'^2(r(t_i),L) + \frac{8}{\Delta t^2} \dot r^2(t_i)),
\end{equation}
where $\Delta t = T / N$. This shows that $\tro$ is non-negative, and that it vanishes 
(as does the second variation $\delta^2 S$) identically when
$V(r)$ exhibit an extremum, and $E$ equals the value of $V(r)$ at
this point, which also implies (from (\ref{eq:rdot})) that $\dot r=0$.
In other words, if the system admits circular orbits, they correspond
to the zeros of $\tro$, and viceversa. Zero-eccentricity orbits in 
the Kepler problem and the light ring in a Schwarzschild spacetime are
example of such special solutions.

\subsection{The nonlinear pendulum}

As another simple example, consider the following
dynamical system:
\begin{equation}
    \label{eq:pendulum-ode}
    \ddot \theta(t) = - \sin \theta(t),
\end{equation}
which describes the evolution of the angle $\theta(t)$ between a pendulum's axis
and the vertical. For $\theta(0)=0$, the solution manifold has two qualitatively
different families of solutions, depending on the initial value of $\dot \theta$, 
or equivalently of the system's energy. For $\dot \theta_0 = \dot \theta(0) < 2$, 
the angle will oscillate between $- \arccos(1-\dot \theta_0^2/2)$ and 
$\arccos(1-\dot \theta_0^2/2)$, whereas for $\dot \theta_0 > 2$ it will grow
monotonically, and the pendulum will reach its apex (at $\theta=\pi$) with 
non-zero speed. The pendulum will therefore complete full revolutions without ever inverting its angular velocity. The separatrix, obtained for  
$\dot \theta_0 = 2$, corresponds to the case where the pendulum has just enough
energy to reach the apex with zero speed, tending to a time-independent end
state given by $\theta=\pi$.

\begin{figure}[!h]
\includegraphics[width=\linewidth]{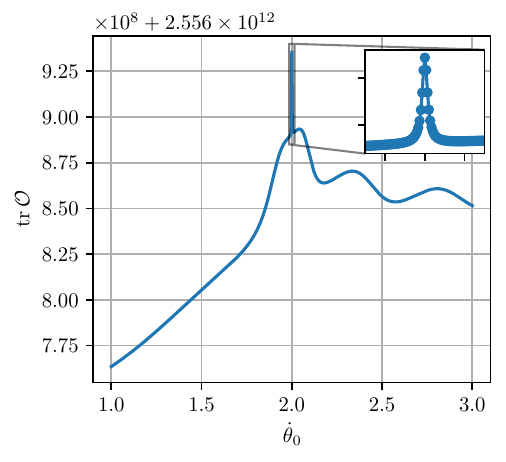}
\caption{Trace of the $\cal{O}$ operator for a family of 
solutions across the transition between back-and-forth oscillations 
($\dot \theta_0 < 2$), and full revolutions ($\dot \theta_0 > 2$).}
\label{fig:pendulum}
\end{figure}

Calculating $\tro$ as a function of $\dot \theta_0$ is straightforward (see Appendix \ref{sec:appB}). We discretise solution trajectories into 1000 points, then calculate the discretised operator $\mathcal{O}$ and its trace using centred finite difference approximations \cite{findiff} (we use the same discretisation for all examples below).
The result is shown in Figure~\ref{fig:pendulum}. We observe a sharp peak
around the transition; the maximum is exactly at the separatrix solution
for $\dot \theta_0 = 2$. Additionally, we observe qualitatively distinct 
regimes on either side of the transition, with $\tro$ taking on
qualitatively distinct values.

A numerical convergence analysis 
is provided in Appendix~\ref{sec:appA}. The expression for $\tro$ is 
shown in Appendix~\ref{sec:appB}. Both are provided for this and all subsequent examples.

\subsection{Population dynamics and the Allee effect}
\begin{figure}
    \includegraphics[width=\linewidth]{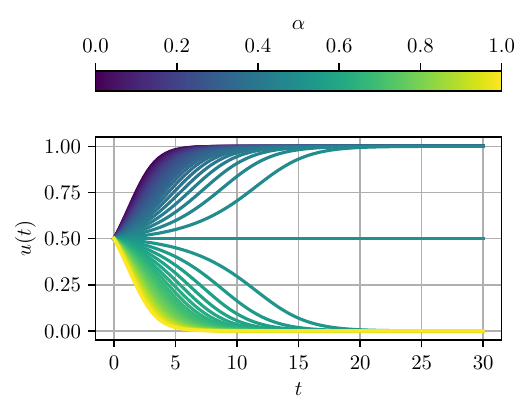}
    \caption{Solutions to (\ref{eqn:firstallee}), describing the single population Allee effect, for different values of the parameter $\alpha$. Trajectories are coloured
    based on the corresponding value of $\alpha$.}
    \label{fig:allee-spatially-homogeneous}
\end{figure}

We next examine an example from population dynamics. We consider a nonlinear ordinary differential equation 
describing the growth of a single population, modulated by the strength of the Allee 
effect~\cite{alma990122856740107026}, defined as a positive association between per capita growth rate and population density~\cite{STEPHENS1999401} and controlled here by the parameter $\alpha$:
\begin{equation}
\label{eqn:firstallee}
    \dot{u} = u(1-u)(u-\alpha).
\end{equation}
This equation describes the evolution of the population density $u(t)$, starting from the initial condition $u(0)=u_0$. It is well known that this model has three steady states $u^* = 0, 1,$ and $\alpha$; the former two are stable, and the latter is unstable. Thus, $\alpha < u_0$ produces growth tending to saturation $u \to 1$ as $t \to \infty$, and $\alpha > u_0$ death, with the population tending to extinction as $u \to 0$ for $t \to \infty$. A special, stationary solution exists for $\alpha = u_0$, the threshold value. Physically, this corresponds to positive feedback of a disadvantage in the population's growth at a lower density. This causes extinction, when below the threshold, and the opposite when above the threshold.
We generate a family of solutions by choosing $u_0 = 0.5$ and integrating (\ref{eqn:firstallee})
in the time interval $[0,30]$, for $\alpha \in [0,1]$ with $\Delta \alpha=0.02$.
Example trajectories are shown in Figure \ref{fig:allee-spatially-homogeneous}.
As expected, the special solution occurs when $\alpha = 0.5$.

\begin{figure}
    \includegraphics[width=\linewidth]{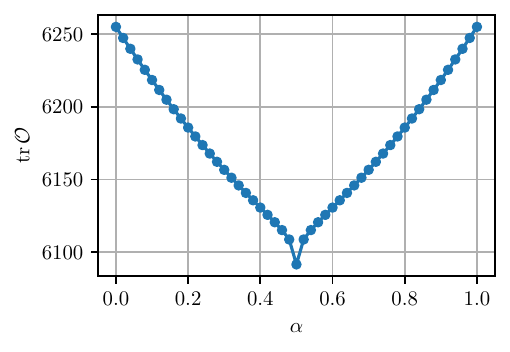}
    \caption{$\tro$ for the single population Allee effect model (\ref{eqn:firstallee}), evaluated for different values of the parameter $\alpha = 0, 0.02, \ldots, 1$.}
    \label{fig:allee0d-trO}
\end{figure}

We can now calculate $\mathcal{O}$ and $\tro$ for each of these solutions through a numerical discretisation with 100 points. The results are shown
in Figure \ref{fig:allee0d-trO}. $\tro$ has a minimum at the critical
value of $\alpha$, meaning that the equation residual is minimally sensitive to solution deformations
for this value of $\alpha$.

\begin{figure}
    \includegraphics[width=\linewidth]{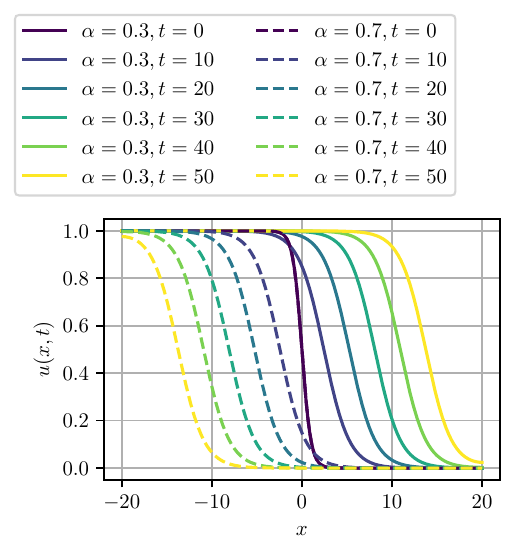}
    \caption{Solution to \eqref{eqn:allee-spatial} at $t= 0, 10, 20, 30, 40, 50$ for $\alpha=0.3$ (solid lines) and $\alpha=0.7$ (dashed lines). The solution asymptotically tends to the shape of an Allee wavefront, with the evolved profiles approximately travelling waves far from the boundary.}
    \label{fig:allee-spatial-solutions}
\end{figure}

\begin{figure}
    \includegraphics[width=\linewidth]{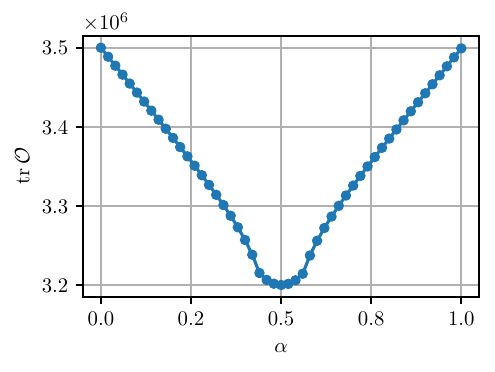}
    \caption{$\tro$ for the spatially inhomogeneous Allee effect model \eqref{eqn:allee-spatial}, evaluated for $\alpha = 0, 0.02, \ldots, 1$.}
    \label{fig:d2S_allee_1d}
\end{figure}

Let us now consider the spatial extension to this problem:
\begin{equation}
    \label{eqn:allee-spatial}
    \frac{\partial u}{\partial t} = \frac{\partial^2 u}{\partial x^2} + u(1-u)(u - \alpha),
\end{equation}
for the scalar field $u(x, t)$, defined on $(x, t) \in [-L, L] \times [0, t_f]$.
For this equation, it is well known that it is possible to make the change of variables $u(x, t) = U(x - ct) = U(z)$, and that such \textit{travelling-wave} solutions exist. Provided we supplement with the boundary conditions $U(- \infty) = 1$ and $U(\infty) = 0$, the wave travels with unique speed $c^* = \sqrt{2} (1 - 2 \alpha) $, describing an invading front when $\alpha < 1/2$, and a receding front when $\alpha > 1/2$ \cite{Hadeler1975}. This corresponds to the balance between the diffusion and reaction terms evaluated on the edge of the wavefront; in physical terms, for $\alpha > 1/2$, diffusion of the population into the space is no longer compensated for by growth of the population in the invading space, leading to recession of the population. In this section, we show that a minima appears in the landscape of $\tro$ at the special solution with zero speed when $\alpha = 1/2$. We generate a family of solutions, starting from the initial condition:
\begin{equation}
    u(x, 0) = \frac{1}{1 + \exp{\left ( \frac{x}{l_i}\right )}},
\end{equation}
and obeying Neumann boundary conditions
\begin{equation}
    \frac{\partial u}{\partial x}(-L, t) = \frac{\partial u}{\partial x}(L, t) = 0.
\end{equation}
Choosing $t_f = 200$, $L = 20$, $l_i = 0.5$, and performing simulations using the finite element method implemented in FENICSX~\cite{Scroggs2022, 10.1145/3524456, baratta_2023_10447666}, we simulate an ensemble of solutions for different values of $\alpha$. Two example solutions demonstrating the possible behaviours are shown in Figure \ref{fig:allee-spatial-solutions}.
From the algebraic expression for $\mathcal{O}$ (see Appendix \ref{sec:appB}), we calculate $\tro$ over the range $\alpha \in [0, 1]$ using 200 time and 80 spatial points. We find that the transition from growth to extinction behaviour is reflected in a minimal value of this quantity (Figure \ref{fig:d2S_allee_1d}). Here, the solution exhibits a shallow minimum.

In the examples presented so far, a single transition in a single scalar field has been responsible for a change in behaviour. We now wish to understand the behaviour of $\tro$ in a model with multiple populations. In particular, we consider a model with Allee kinetics in the prey and saturating predation in the absence of diffusion; including diffusivities for the species has been studied previously~\cite{petrovskii_regimes_2005}. For the population densities of the prey $u$ and predator $v$, the model equations read
\begin{align}
    \label{eqn:predprey_allee1}
    \dot{u} &= \gamma u(1-u)(u-\beta) - \frac{uv}{1+ \alpha u}, \\
    \label{eqn:predprey_allee2}
    \dot{v} &= \frac{uv}{1 + \alpha u} - \delta v,
\end{align}
with $\alpha, \beta, \gamma, \delta$ as positive constants. This system describes a prey population, $u$, that needs to be over a threshold density to grow, and which is predated upon by a predator population, $v$, with maximal per capita rate $1$. The predator has constant per capita death rate $\delta$. This model has four steady states,
\begin{align}
    (u^*, v^*) &= (0, 0), \\
    (u^*, v^*) &= (1, 0), \\
    (u^*, v^*) &= (\beta, 0), \\
    (u^*, v^*) &= \left ( \frac{\delta}{(1 - \alpha \delta)}, \frac{\gamma(1 - \delta(\alpha+1))(\delta (\alpha \beta + 1) - \beta) }{(1 - \alpha \delta)^3} \right ).
\end{align}
Phase plane analysis of this system can show that if we enforce existence of a \textit{coexistence steady state} (a steady state with both $u^*, v^* > 0$; physicality demands that neither can be negative), then the appropriate parameter range to consider is
\begin{align}
    \beta &< 1, \\
     (\alpha + \beta^{-1})^{-1} &< \delta < (1 + \alpha)^{-1}. \label{eqn:coexistence-condition-2}
\end{align}
Outside of this range, $v^* \leq 0$ for all four steady states. Whilst the coexistence state exists, the steady state at $(u^*, v^*) = (0, 0)$ is stable, and the rest are saddle points.
The question we examine is under what conditions (on $\beta$ and $\delta$) the supplied initial conditions $(u(0), v(0))$ will lie within the basin of attraction of the coexistence state (so that coexistence will be the end state) and when, viceversa, the end state will be one of the three $v^*=0$ fixed points. This dual behaviour is an example of a global bifurcation, and the
question cannot be answered with a simple local stability analysis. This example highlights the power of the 
$\tro$ score.

\begin{figure}
    \includegraphics{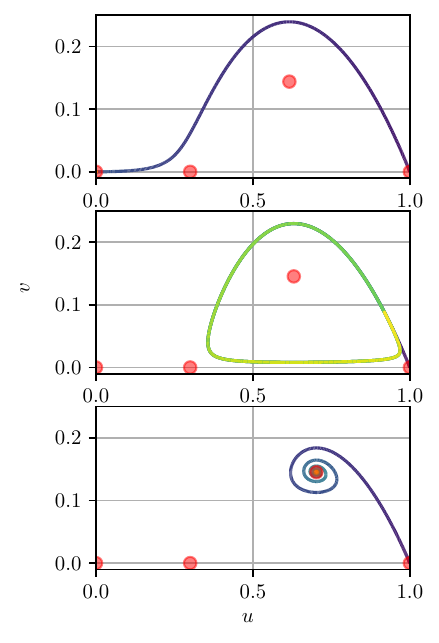}
    \caption{Sample realisations of the predator-prey model \eqref{eqn:predprey_allee1}, \eqref{eqn:predprey_allee2} for parameters $\beta=0.3$, and varying $\delta=0.52, 0.53$ and $0.58$ (top to bottom, respectively). The four fixed points are plotted in red. Trajectories are coloured by simulation time, where navy indicates the start of the simulation, and yellow indicates the end.}
    \label{fig:predator_prey_sample_orbits}
\end{figure}

In what follows, we make the choice that $\gamma = 1, \alpha=0.3, u(0) = 1, v(0) = 0.001$, and consider only the effect of changing the parameters $\beta$ and $\delta$. Simulation results in the $(u, v)$ phase plane are shown in Figure~\ref{fig:predator_prey_sample_orbits} for sample values of $\delta$ holding $\beta$ constant. 

\begin{figure}
    \includegraphics[width=\linewidth]{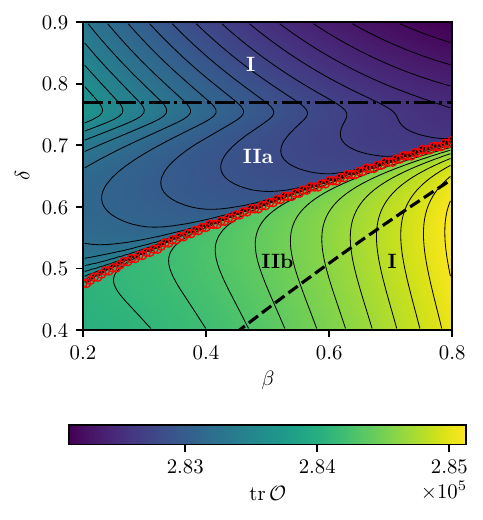}
    \caption{Calculation of $\tro$ for the predator-prey system using a high-resolution sweep in parameter space, with $\alpha=0.3$ and $\gamma=1$. In region I, all of the system's physically relevant fixed points lie on the $u$
    axis. In region IIb, a fourth fixed point with positive $v$ appears, but is not reached starting from the initial 
    conditions $u(0) = 1, v(0) = 0.001$. In region IIa, the fourth fixed point exists and is the end point of the trajectories
    that start at $u(0),v(0)$. The red empty circles represent the points where the transition between IIa and IIb is 
    observed in the numerical integration of \eqref{eqn:predprey_allee1},\eqref{eqn:predprey_allee2}.}
    \label{fig:allee-db}
\end{figure}

Varying the value of $\delta$ allows the system to exhibit a range of behaviours. For sufficiently high predator death rate, $\delta$, the system first shows coexistence (in the form of a small amplitude limit cycle relative to the predator and prey population densities), where the predator death rate is too fast relative to the prey growth rate. This prevents the prey from approximately reaching a steady state density, without the predator also approaching a steady state density. Decreasing $\delta$ then allows the growth of the predator population to lag that of the prey population, producing large amplitude limit cycle oscillations in the two populations. Decreasing $\delta$ further then allows predation to overpower the prey's growth and bring its density below the threshold value $\beta$, leading to mutual extinction. The relevant region of the $(\beta,\delta)$ plane can then
be subdivided into three regions, as shown in Figure~\ref{fig:allee-db}: region I, where only the three $v^*=0$ fixed points
exist and defined by \eqref{eqn:coexistence-condition-2}; region IIa, where a fourth fixed point with positive $v^*$ appears, and is the end state of trajectories starting
from the prescribed initial conditions; and region IIb, where the fourth fixed point exists but is unreachable.

Our simulated results are highly suggestive of a heteroclinic cycle which vanishes below a certain $\delta$ for given $\beta$. 
To study this behavior, we perform an ensemble of simulations with $\mathrm{\Delta}\beta = 0.006,$ $\mathrm{\Delta}\delta = 0.005$ in the range $(\beta, \delta) \in [0.2, 0.8] \times [0.4, 0.9]$.
We calculate $\tro$ through a discretisation with 100 points over the $(\beta, \delta)$ plane we obtain the profile shown in Figure~\ref{fig:allee-db}. The most
prominent feature is a sharp decrease of this quantity over the separatrix between regions IIa and IIb. We note that $\tro$
is also sensitive to other transition regions, such as the line between I and IIa, where $\tro$ for fixed $\beta$ has a local
maximum. This leads to a key realisation: Detecting transitions in multidimensional parameter spaces will generally be 
more complex than transitions governed by a single parameter, where a zero or a stationary point in $\tro$ may be the only 
diagnostics of a heightened solution sensitivity. In multidimensional spaces, the interplay between the solution variability
along different directions may create a complex response in $\tro$.

Also notice that, as in previous examples in this section, there is no dramatic change in the solutions themselves. 
The system simply transitions from a regular behaviour (a transitory predator-prey interaction followed by mutual extinction)
to another regular, albeit qualitatively different, one (self-sustaining coexistence of the predator and prey populations).
The role of $\tro$ is thus to provide a measure of qualitative transitions, which could be used for searching a solution
space via traditional optimisation methods. 

\subsection{The Amari neural model}

Dynamical systems are frequently used in neural models, where brain activity is examined via the 
evolution and coupling of different neuronal populations. An important example is the Amari model,
which has been previously used to investigate the mechanisms of Epileptic spike-wave (SW) seizures \cite{taylor2014computational}.
SW seizures can be interpreted as a dynamical phase in a neuronal
system~\cite{da2003epilepsies}. 
The SW seizure phase and the background (non-seizure) phase coexist,  
are locally stable, and therefore transitions between the two can be triggered by stimulation~\cite{suffczynski2004dynamics}.

The model describes brain activity as the inhibitory/stimulatory interplay 
between four different populations:
cortical pyramidal neurons ($P$), cortical inhibitory interneuron ($I$), thalamo-cortical neurons ($T$), and inhibitory thalamic reticular neurons ($R$).
The interplay is represented by the following coupled ODEs:
\begin{eqnarray}
    \dot P &=& t_1 (h_P - P + c_1 f(P) - c_3 f(I) + c_9 f(T)), \label{eq:a1}\\
    \dot I &=& t_2 (h_I - I + c_2 f(P)), \label{eq:a2} \\
    \dot T &=& t_3 (h_T - T + c_7 f(P) - c_6 s(R)), \label{eq:a3} \\
    \dot R &=& t_4 (h_R - R + c_8 f(P) - c_4 s(R) + c_5 s(T)), \label{eq:a4}
\end{eqnarray}
where
\begin{eqnarray}
    f(u) &=& ( 1 + \epsilon^{-u} )^{-1}, \label{eq:amari-sigmoid} \\
    s(u) &=& a u + b, \label{eq:amari-linear}
\end{eqnarray}
are activation functions and the system's parameters are defined in Appendix~\ref{app:amari}.

\begin{figure}
    \centering
    \includegraphics{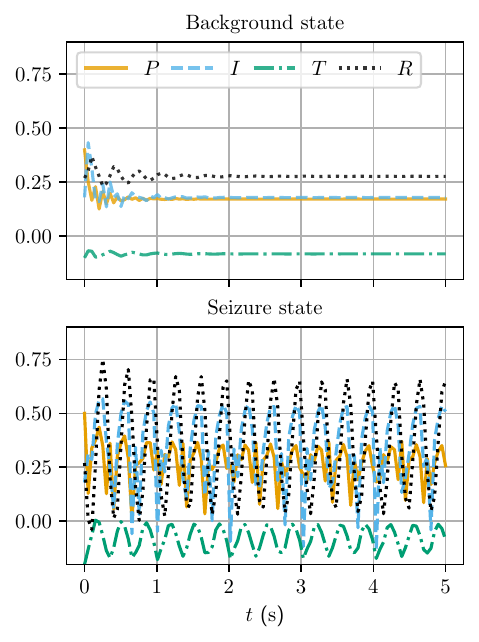}
    \caption{Example trajectories of the system (\ref{eq:a1})-(\ref{eq:a4}), for $(P_i,I_i,T_i,R_i)=(0.4,0.18,-0.1,0.27)$ (top, background state) and
    $(P_i,I_i,T_i,R_i)=(0.4,0.18,0,0.27)$ (bottom, seizure).}
    \label{fig:nstates}
\end{figure}

This model can be used to classify which initial conditions on $(P,I,T,R)$ evolve
into a seizure state. Two solutions to this system, one representing a seizure (characterised by strongly oscillatory behaviour) and
one converging to the background state (where the four dynamical variables approach
steady-state values), are shown in Figure~\ref{fig:nstates} along
with the corresponding initial conditions.

Identifying the basin of attraction of the background state transcends pure academic 
interest. Emerging techniques like neurostimulation~\cite{starnes2019review} provide an effective method
to treat seizures, which may be enhanced if the required stimulation to nudge the system into
the basin of attraction of the background state can be computed accurately and efficiently.
This requires the identifications of the boundaries of this domain in a multidimensional
space.

\begin{figure}
    \centering
    \includegraphics[width=\linewidth]{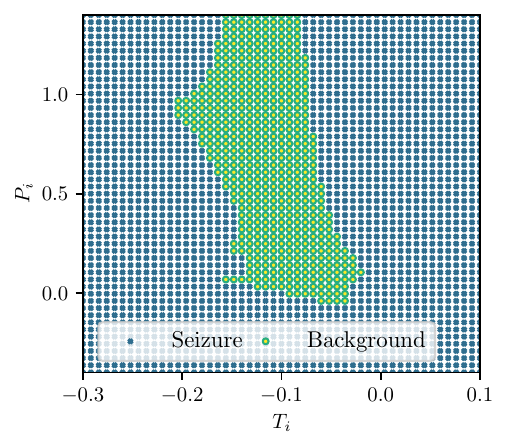}
    \caption{Classification of trajectories of the Amari system in the $(T_i,P_i)$ plane for $I_i=0.18$, $R_i=0.27$.}
    \label{fig:class}
\end{figure}

The problem can be tackled with a brute-force approach, where the parameter space is swept
along all the relevant directions, and each resulting trajectory is classified. For example,
we can classify 
a set of initial conditions as ending within the background state, if after 3 seconds 
the model output does not exceed a suitably chosen threshold \cite{taylor2014computational}.
For ease of visualisation, we hold two of the initial conditions fixed ($I_i=0.18$ and $R_i=0.27$)
and show the result of the classification in Figure~\ref{fig:class}.

This procedure, however, rapidly becomes computationally intensive when the search
is performed in a higher-dimensional space (to explore the full set of initial conditions,
or extend to models with different values of the configuration parameters in Appendix \ref{app:amari}).
As for other systems, we show that $\tro$ provides a convenient score to guide a search
of this parameter space.

\begin{figure}[!h]
    \centering
    \includegraphics[width=\linewidth]{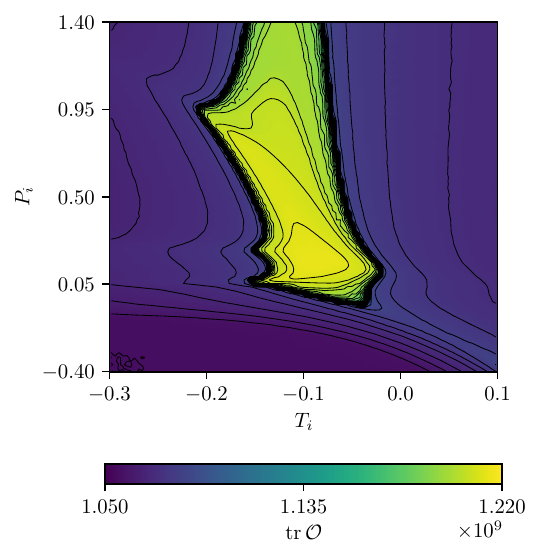}
    \caption{Value of $\tro$ of the Amari system in the $(T_i,P_i)$ plane.}
    \label{fig:amariO}
\end{figure}

To illustrate this, we first compute the $\mathcal{O}$ operator from the system (\ref{eq:a1})-(\ref{eq:a4})
for a family of 2500 trajectories, obtained setting $(I_i,R_i)=(0.18,0.27)$ and 
varying $T_i=[-0.3,0.1]$ and $P_i=[-0.3,1.4]$. Each trajectory is computed for
$t=[0,5]$ and discretised into $N=100$ points. $\tro$ can then be found from $\mathcal{O}$, similarly to previous examples. Its numerical value for this family of solutions is shown in Figure~\ref{fig:amariO}.

Two aspects are evident: Firstly, the values of $\tro$ change sharply at the 
boundary of the attraction basin, providing an independent metric to distinguish the two dynamical phases. Secondly, its contour
lines are sensitive to the existence of this transition well outside of its 
immediate neighborhood, so that ascending or descending the gradient of $\tro$
is an effective guide towards the discovery of either class. Once a portion of 
the boundary is discovered, techniques such as nested sampling or edge tracking can
be used to continue it with a minimal amount of model evaluations.

\subsection{Light propagation in a gravitational field}
\label{subsec:JW-Sachs_Equations}

Next, we turn our attention to the Sachs equations \cite{Sachs61}, a system of four coupled nonlinear ODEs 
representing the evolution of light bundles as they propagate through a gravitational
field. Light phenomenology on curved space is extremely rich, admitting many unique
solutions like Einstein rings of galaxies \cite{Einstein36} and other gravitational-lensing
behavior.
Analyses of the lensing effect and the light deflected from a range of astronomical sources can be used as powerful probes of cosmology and fundamental physics \cite{Bartelmann10}.

For a general space time metric and stress-energy tensor, the Sachs equations are given by \cite{Yoo_cosmo}:
\begin{align}
    \label{eq:JW-theta}
    & \dot \theta+\frac{1}{2}\theta^{2}(\lambda)+2\sigma^{2}(\lambda) -2\omega^{2}(\lambda)+ \mathcal{R}=0 \\
    \label{eq:JW-omega}
    & \dot \omega +\theta(\lambda)\,\omega(\lambda)=0 \\
    \label{eq:JW-sigma_plus}
    & \dot \sigma_{+}+\theta(\lambda)\sigma_{+}(\lambda)+\mathcal{C}_{+}=0 \\
    \label{eq:JW-sigma_times}
    & \dot \sigma_{\times}+\theta(\lambda)\sigma_{\times}(\lambda)+\mathcal{C}_{\times}=0
\end{align}
where a dot denotes differentiation with respect to $\lambda$ (an affine parameter that can be interpreted as a distance, or time measure along the path of propagation), 
and $\theta$, $\omega$, and $\sigma^2 ={\sigma_+}^2 + {\sigma_\times}^2$ are the Sachs
optical scalars, representing, respectively, the expansion, rotational twist, and shear
the beam's cross sectional area undergoes as it propagates along a path in spacetime. $\mathcal{R}$, $\mathcal{C}_{+}$, and 
$\mathcal{C}_{\times}$ are scalars representing the curvature of the
underlying spacetime, so that the Sachs system effectively models how gravitational
fields bend light.

In this section, we will study this system when the spacetime is described
by the de-Sitter model, a spatially flat Friedmann-Lema{\^i}tre-Robertson-Walker cosmology \cite{Dodelson20} with $\mathcal{R}={\cal{C_{+}}}={\cal{C_{\times}}}=0$.

Despite the model's simplicity, light beams can follow distinctly different evolution 
paths depending on the initial values of the expansion, rotation, and shear scalars.
In particular, certain initial conditions $(\theta_0, \omega_0, \sigma_0)$
will lead to optical singularities, where one or more of the scalars
blow up and the beam's cross-sectional area collapses to zero. Singular trajectories are obtained, provided that $\sigma_0^2 \geq \omega_0^2$, and either (see Appendix~\ref{app:sachs})
\begin{itemize}
    \item $\theta_0 \leq 0$, or
    \item $\theta_0 > 0$ and $\theta_0^2 < 4(\sigma_0^2 - \omega_0^2)$.
\end{itemize}

\begin{figure}
\centering
\includegraphics[width=\linewidth]{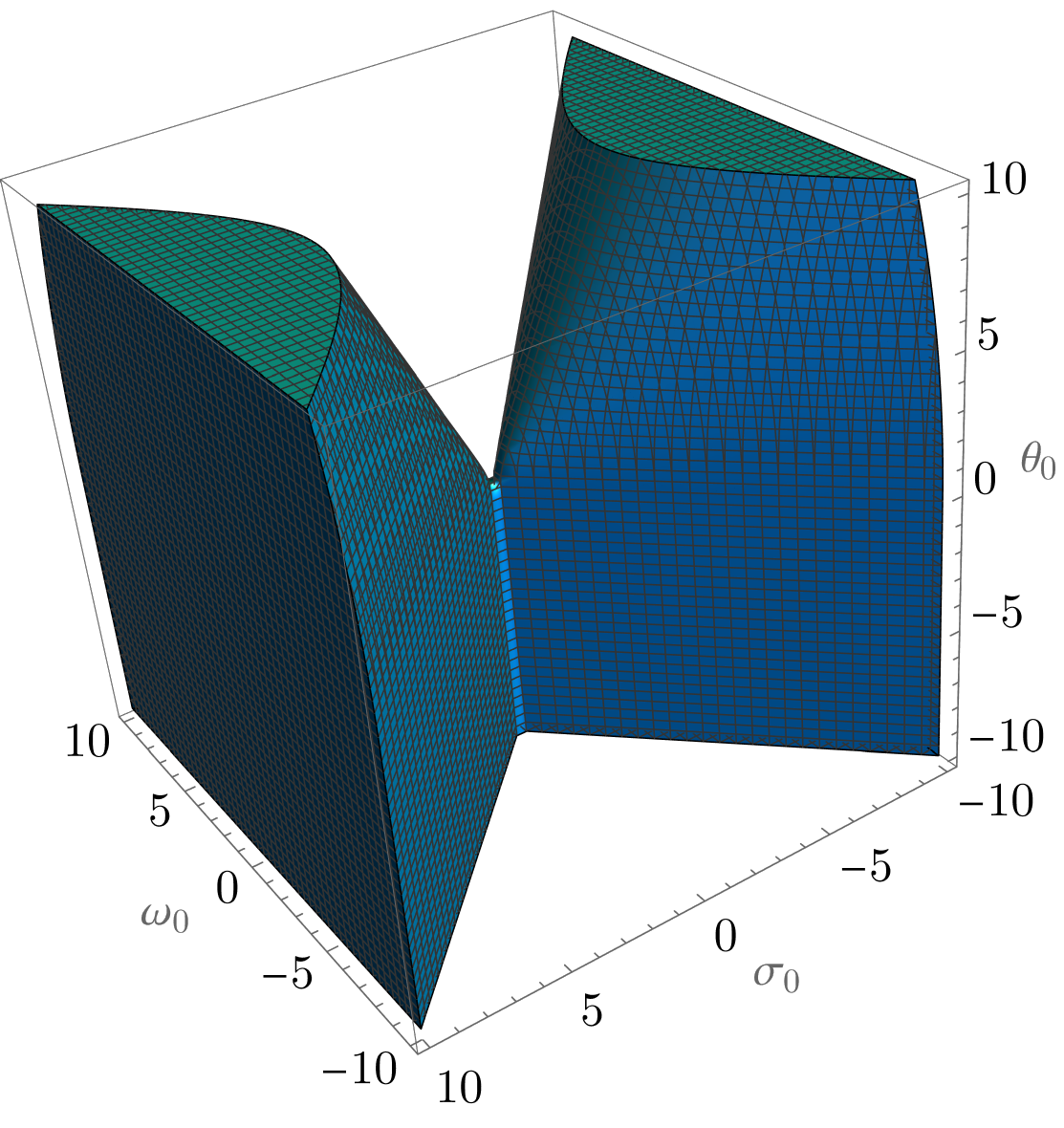}
\caption{\label{fig:surf}Regions in the initial-condition space corresponding
to singular trajectories (enclosed by the blue surface around the $\omega_0=0$ axis)
and regular ones (outside of the surface).}
\end{figure}

The surface between singular and regular solutions is shown in Figure~\ref{fig:surf}. Notice that this is the only case considered 
here where some solutions are not regular. 

\begin{figure}
\centering
\includegraphics[]{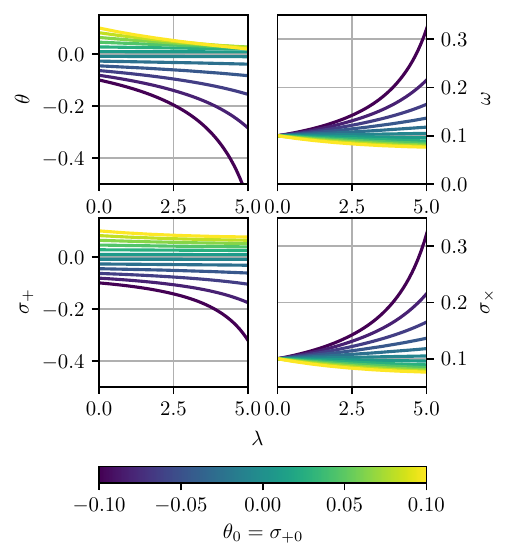}
\caption{\label{fig:traj-to}A family of curves obtain setting $\omega_{0}=\sigma_{\times,0}=0.1$ and
$\theta_0$ and $\omega_0$ both varying in $[-0.1,0.1]$, respectively.}
\end{figure}

\begin{figure}
\centering
\includegraphics[]{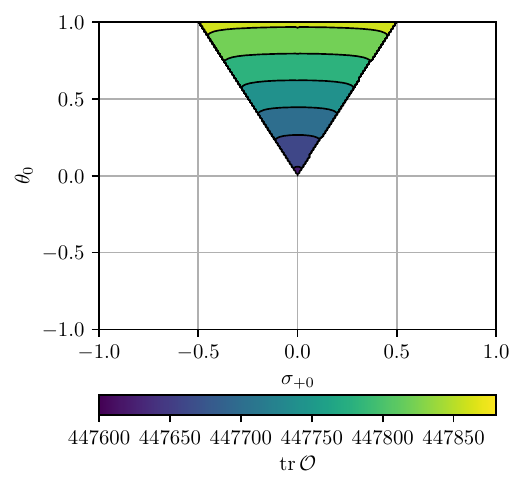}
\includegraphics[]{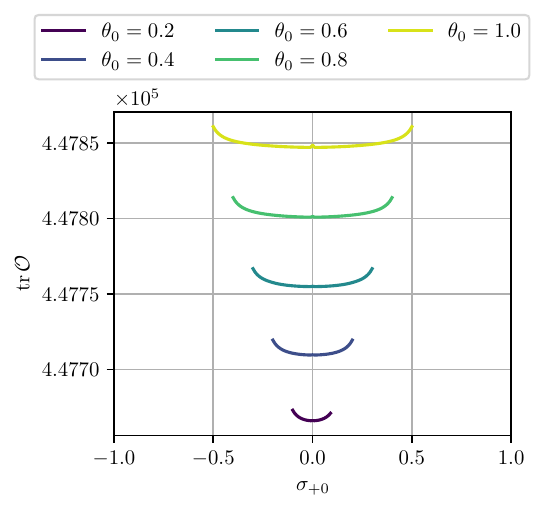}
\caption{\label{fig:tro-sachs}The behaviour of $\tro$ for the Sachs system
on the $\theta_0$-$\sigma_{+0}$ plane. Outside of the shown region, the 
system behaviour is singular and $\tro$ is not shown.}
\end{figure}

We can construct $\cal{O}$ for this system, and observe how $\tro$ behaves across this surface.
To explore this transition, we numerically evolve a family of solutions of the Sachs system 
for $\lambda \in [0,5]$
and different initial conditions $\theta_{0}, \omega_{0}, \sigma_{+0}, \sigma_{\times,0}$, with 200 points.
Example solutions are shown in Figure~\ref{fig:traj-to}. This family can be used to compute
the $\tro$ score, as shown in Figure \ref{fig:tro-sachs}.

We examine the score in the region where
the trajectories, and therefore $\tro$, are not singular. For constant 
$\theta_0$, $\tro$ exhibits little variation away from the transition
points $\sigma_{+0} = \pm \theta_0/2$; at these boundaries, on the other
hand, the score increases sharply. For $\sigma_{+0}=0$, the score
decreases monotonically towards the origin, where the system's behaviour 
again changes to singular.

We note that these transitions are more difficult to identify than in previous
cases. It is possible that $\tro$ is less suitable to explore 
transitions to singular dynamics, that it is more difficult to compute accurately 
around singularities (but see the error estimation section below), or that more 
sophisticated optimisation may be required in those cases.

\section{Conclusions}
\label{sec:concl}
We have discussed the challenge of discovering dynamical transitions
in systems whose behavior is governed by one or more nonlinear 
differential equations. We have presented an argument relating such
transitions to the sensitivity of the governing equations' residual
on arbitrary solution deformations. This sensitivity can be 
expressed as the trace of the operator $\tro$ appearing in the second
variation of the residual's $L_2$ norm under such deformations.

We showed that, as a functional of the solution $u$, this sensitivity
is only nontrivial for nonlinear systems, and discussed additional universal properties of this quantity.
Next, we have illustrated our approach on known dynamical systems and partial differential 
equations across a range of domains. We showed that changes parameterised by initial and boundary 
conditions, and parameters within the residual showed features (such as extrema) in the score landscape. Crucially, 
we emphasise that features in the score captured global bifurcations,
which elude local-stability analyses and
can generally only be discovered via trial and error. 
Unlike approaches aimed at detecting quantitative dynamical changes, the score proposed here 
can capture qualitative changes in behaviour in an application-agnostic manner, even without 
a quantitative anomaly. The main objective of this
work is to report and illustrate this property across different 
applications; we reserve explaining the full phenomenology of $\tro$
for a future study.

Finally, we have discussed a number of limitations of this approach.
Firstly, we have shown that, unlike the residual itself, $\tro$ is not 
guaranteed to remain regular for singular solutions.
This complicates its use in cases where the solution or any of its derivatives 
blow up in finite time.

Secondly, all dynamical shifts discussed here are well known and easily computable;
whether the score proposed here is viable to
calculate and optimise in much higher-dimensional cases, and can lead
to genuine discovery of previously unknown dynamical regimes, remains to be determined. Initial characterisation comparing $\tro$ to other measures of solution diversity in Appendix \ref{sec:tro-sensitivity-evaluation} suggests that $\tro$ shares information with other approaches, but may contain additional information which could be leveraged for optimisation. We will address these limitations in future studies.

Code for the computation of the operator $\cal{O}$ and its trace for 
arbitrary differential equations is in preparation for release.

\section{Acknowledgements}
The authors are indebted to Tobias Grafke, Valerio Lucarini,
Peter Taylor, and Yujiang Wang for 
invaluable conversations and suggestions for improvement. All 
authors acknowledge support from a UKRI Future Leaders Fellowship
(MR/T041862/1).

\appendix

\section{Numerical error analysis}
\label{sec:appA}

In all cases considered, we have computed the trace of the operator $\cal{O}$ after 
discretization. In this section, we analyze the convergence of this procedure in the
continuum limit, i.e.~for increasing mesh sizes. Notice that the convergence of the
trace is not guaranteed, as only trace-class operators admit a finite limit. However,
divergences can be renormalized away in practice (see e.g.~the discussion 
in~\cite{2022arXiv221109476G} and references therein). Here, we predict the overall scale based on the 
finite-difference representation of the differential operators contained in $\cal{O}$,
and divide the trace by this quantity to obtain a convergent series.

We illustrate this procedure in this section, and note that the interesting properties
of $\tro$ (such as transitions and their parametric location) are unaffected by this
particular choice of renormalization.

To predict the dependence of $\tro$ on the number of points $N$ used to discretize the
fields and their derivatives, we observe that for a differential equation of order $n$,
the operator $\cal{M}$ will contain derivatives up to order $2n$, resulting in
terms proportional to $\delta^{-2n}$ in the discretization of $\cal{O}$, where $\delta$ is a
measure of the spatiotemporal spacing of the numerical mesh (in multidimensional
cases, we assume that the convergence study is performed by decreasing the 
spacing in all dimensions simultaneously). When computing the trace, the number of 
elements included in the sum will scale as $\delta^{-1}$, resulting in an overall
factor of $\delta^{-(2n+1)}$, or $N^{2n+1}$.

\begin{figure}
    \centering
    \includegraphics[width=0.9\linewidth]{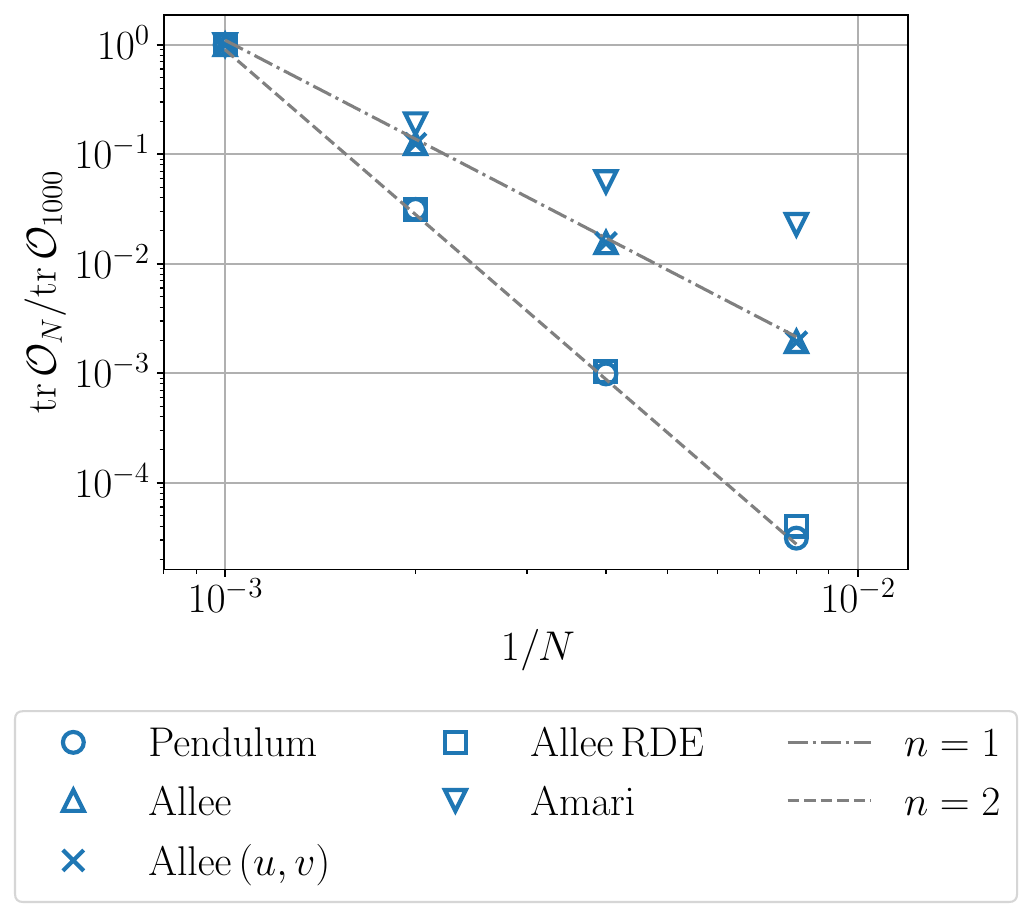}
    \caption{Scale of $\tro$ as a function of the number of discretization points $N$.}
    \label{fig:trconv}
\end{figure}

Figure \ref{fig:trconv} shows the dependence of $\tro$ (averaged over the parametric interval
considered) on $N$ for all examples discussed, with $N = [125, 250, 500, 1000]$. The measured trends correspond, in all cases,
to the expected convergence order, based on the highest derivative order present in the
equations. The agreement with the expected convergence is affected, in practice, by 
how much the $N^{(2n+1)}$ term dominates over the lower-order terms in the interval
$[125, 1000]$. The Amari case, where the constant coefficient is much larger than the 
coefficient of the $N^{3}$ term, shows the largest deviation from the asymptotic
behaviour in this interval.

\begin{figure*}
    \centering
    \includegraphics[]{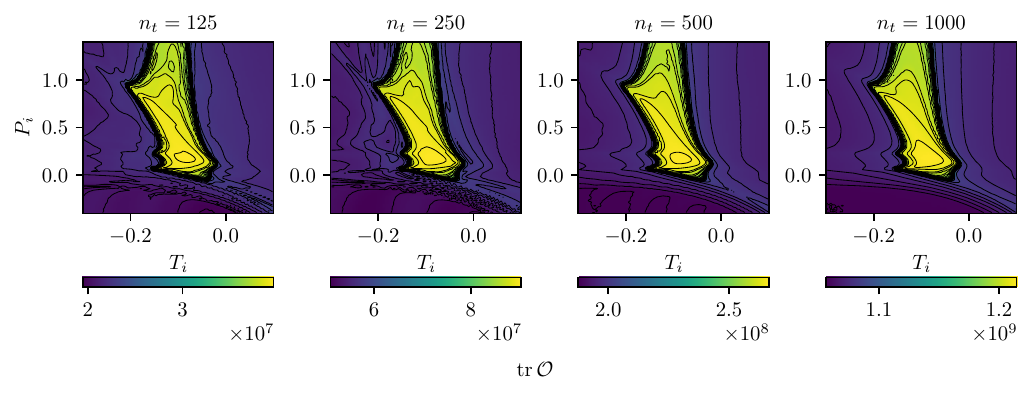}
    \caption{Parametric dependence of landscape of $\tro$ at different discretisation resolutions $n_t=125, 250, 500, 1000$ whilst holding the total simulation time fixed.}
    \label{fig:amari-convergence}
\end{figure*}

The rescaled $\tro/N^{(2n+1)}$ can then be used to ensure that the numerical error
decreases in the limit of increasing resolution. Crucially, in addition to the overall scale, 
we have also examined the profile of this quantity over the region of solution 
space explored, and found that the corresponding optimisation landscape is insensitive
to $N$. Features such as extrema and step changes occur at the same parameter values
regardless of resolution. Figure \ref{fig:amari-convergence} demonstrates this property for the Amari case.

\section{Explicit expressions for $\mathcal{O}$}
\label{sec:appB}
In the main text, we showed that the second variation of $S[u + \eta]$ can be rewritten as
\begin{equation}
    \delta^2 S = \int \eta^{\top} \mathcal{O} \eta \; {\rm d} t,
\end{equation}
where $\mathcal{O}$ is a differential operator evaluated using a solution of the given differential equation(s). Below, we provide explicit expressions for $\mathcal{O}$ for the examples presented.
\subsubsection*{The nonlinear pendulum}
For the nonlinear pendulum, we can show that
\begin{equation}
    \mathcal{O} = 2{\rm d}_{tttt} + (4 \cos{\theta}) {\rm d}_{tt} + 2(\cos^2{\theta} - \sin^2{\theta} - \ddot{\theta}\sin{\theta}).
\end{equation}
\subsubsection*{The Allee effect models}
For the single-species Allee effect \eqref{eqn:firstallee}, it can be shown by applying \eqref{eq:O-general-expression} that
\begin{equation}
    \mathcal{O} = 2 \left [f_u^2 + (f-\dot u)f_{uu} \right] - 4 f_u {\rm d}_t - 2 {\rm d}_{tt}
\end{equation}
where $f=u(1-u)(u-\alpha)$ and the $u$ subscript denotes differentiation with respect to $u$. For the Allee reaction-diffusion equation \eqref{eqn:allee-spatial}, it can be shown similarly by  evaluating \eqref{eq:O-general-expression} that
\begin{equation}
    \delta^2 S = \int\limits_0^{t_f} \int\limits_{-L}^{L} \eta^{\top}\mathcal{O} \eta \, \mathrm{d}t \mathrm{d}x
\end{equation}
with:
\begin{eqnarray}
    \mathcal{O} &=& 2 \left [f_u^2 + (f-\dot u + u'')f_{uu} \right] \nonumber \\ 
    && - 4 f_u \partial_t
    - 2 \partial_{tt} 
     + 4 f_u \partial_{xx} \nonumber \\
    &&+ 2 \, \partial_{xxxx} 
    + 4 \, \partial_{txx}
\end{eqnarray}
where $f(u) = u(1-u)(u-\alpha)$, and primes or dots denote a partial derivative with respect to $x$ or $t$, respectively.

For the Allee predator-prey model, define $f(u) = \gamma u(1-u)(u-\beta)$ and $g(u, v) = uv/(1+\alpha u)$. We construct $\mathcal{O}$ by first constructing the four 'blocks' corresponding to $\mathcal{O}_{uu}$ (which produces terms only containing perturbations to $u$, $\eta_u$), $\mathcal{O}_{uv}$ (terms which act on perturbations to $v$ but produce mixed terms), $\mathcal{O}_{vv}$, and $\mathcal{O}_{vu}$ which are defined analogously. Then
\begin{align}
    \begin{autobreak}
    \MoveEqLeft
        \mathcal{O}_{uu} = 2((f_{u})^2 
        + ff_{uu} 
        - 2f_ug_u 
        - f_{uu} g 
        - f_{uu} \dot{u} 
        + g_{uu} \dot{u} 
        + 2 g g_{uu} 
        + 2(g_u)^2 
        - g_{uu} \dot{v} 
        - \delta v g_{uu}) 
        - 4(f_u - g_u) {\rm d}_t - 2 {\rm d}_{tt}
    \end{autobreak}
    
    \\
    \begin{autobreak}
    \MoveEqLeft
        \mathcal{O}_{uv} = 2(2(g_u g_v + g g_{uv})
        + \dot{u}g_{uv} 
        - f_u g_v 
        - f g_{uv} 
        - \dot{v} g_{uv} - \delta (g_u + v g_{uv}))
        - 4 g_u {\rm d}_t
    \end{autobreak}
    
    \\
    \begin{autobreak}
    \MoveEqLeft
        \mathcal{O}_{vu} = 2(2(g_u g_v + g g_{uv})
        + \dot{u}g_{uv} 
        - f_u g_v 
        - f g_{uv} 
        - \dot{v} g_{uv} - \delta (g_u+v g_{uv}))
    \end{autobreak}

    \\
    \begin{autobreak}
    \MoveEqLeft
        \mathcal{O}_{vv} = 2(2((g_v)^2 + g g_v) 
        + \dot{u} g_{vv}
        - f g_{v}
        + \delta^2
        - \dot{v} g_{vv}
        - 2\delta (2 g_{v} - v g_{vv}))
        + 4(\delta - g_v) {\rm d}_t
        - 2{\rm d}_{tt}
    \end{autobreak}
    
\end{align}
The full operator is provided by the construction
\begin{equation}
    \label{eqn:Multi-equation-operator-definition}
    \mathcal{O} = \sum_{\{ X_j\}}\sum_{\{ X_k\}} \mathcal{O}_{X_j X_k},
\end{equation}
where for this example $\{ X_i \} = \{u, v\}$.

\subsubsection*{The Amari model}
The Amari model is constructed analogously from $\mathcal{O}_{X_j X_k},$ where $X_j \in \{P, T, I, R\}$. Define the right hand sides of each equation as $\tilde{P}, \tilde{I}, \tilde{T}, \tilde{R}$, respectively. Then 
\begin{align}
    
    \begin{autobreak}
    \MoveEqLeft
    \mathcal{O}_{PP} = 2 ((\tilde{P}_P)^2 + (\tilde{P} - \dot{P})\tilde{P}_{PP}) 
    - 2 {\rm d}_{tt} 
    - 4\tilde{P}_P {\rm d}_t
    + 2 ((\tilde{I}_P)^2 + (\tilde{I} - \dot{I})\tilde{I}_{PP})
    + 2 ((\tilde{T}_{P})^2 + (\tilde{T} - \dot{T}) + \tilde{T}_{PP})
    + 2 \tilde{R}^2 
    + (R - \dot{R})\tilde{R}_{PP}
    \end{autobreak}

    \\
    \begin{autobreak}
    \MoveEqLeft
    \mathcal{O}_{II} = 2 (\tilde{P}_{I}^2 + (\tilde{P} - \dot{P})\tilde{P}_{II})
    + 2(\tilde{I}_{I}^2 + (\tilde{I} - \dot{I})\tilde{I}_{II})
    - 4\tilde{I}_{I} {\rm d}_t
    -2 {\rm d}_{tt}
    \end{autobreak}

    \\
    \begin{autobreak}
    \MoveEqLeft
    \mathcal{O}_{TT} = 2 ( (\tilde{P}_{T})^2 + (\tilde{P} - \dot{P})\tilde{P}_{TT})
    + 2 ((\tilde{T}_{T})^2 + (\tilde{T} - \dot{T}) \tilde{T}_{TT})
    - 4 \tilde{T}_{T} {\rm d}_t
    - 2 {\rm d}_{tt}
    + 2 \tilde{R}^2 + (R - \dot{R})\tilde{R}_{TT}
    \end{autobreak}

    \\
    \begin{autobreak}
    \MoveEqLeft
    \mathcal{O}_{RR} = 2 ((\tilde{T}_{R})^2 + (\tilde{T} - \dot{T}) \tilde{T}_{RR})
    + 2 ((\tilde{R}_{R})^2  + (\tilde{R} - \dot{R})\tilde{R}_{RR} )
    - 4 \tilde{R}_{R} {\rm d}_t
    - 2 {\rm d}_{tt}
    
    \end{autobreak}

    \\
    \begin{autobreak}
    \MoveEqLeft
    \mathcal{O}_{PI} = -4 \tilde{P}_I {\rm d}_t + 2 \tilde{P}_P \tilde{P}_I
    + 2 \tilde{I}_I \tilde{I}_P - 4 \tilde{I}_P {\rm d}_t
    \end{autobreak}

    \\
    \begin{autobreak}
    \MoveEqLeft
    \mathcal{O}_{IP} = 2 \tilde{P}_{P} \tilde{P}_I + 2 \tilde{I}_I \tilde{I}_P
    \end{autobreak}

    \\
    \begin{autobreak}
    \MoveEqLeft
    \mathcal{O}_{PT} = -4\tilde{P}_T {\rm d}_t + 2\tilde{P}_P \tilde{P}_T 
    + 2 \tilde{T}_P \tilde{T}_T
    + 2 \tilde{R}_P \tilde{R}_T
    \end{autobreak}

    \\
    \begin{autobreak}
    \MoveEqLeft
    \mathcal{O}_{TP} = 2 \tilde{P}_P \tilde{P}_T 
    - 4 \tilde{T}_P {\rm d}_t 
    + 2 \tilde{T}_P \tilde{T}_T
    + 2 \tilde{R}_P \tilde{R}_T
    \end{autobreak}

    \\
    \begin{autobreak}
    \MoveEqLeft
    \mathcal{O}_{PR} = 2 \tilde{T}_R \tilde{T}_P
    + 2 \tilde{R}_P \tilde{R}_R
    \end{autobreak}

    \\
    \begin{autobreak}
    \MoveEqLeft
    \mathcal{O}_{RP} = 2 \tilde{T}_R \tilde{T}_P
    - 4 \tilde{R}_P {\rm d}_t 
    + 2 \tilde{R}_{R} \tilde{R}_P
    \end{autobreak}

    \\
    \begin{autobreak}
    \MoveEqLeft
    \mathcal{O}_{IT} = 2 \tilde{P}_{T} \tilde{P}_I
    \end{autobreak}

    \\
    \begin{autobreak}
    \MoveEqLeft
    \mathcal{O}_{TI} = 2 \tilde{P}_{T} \tilde{P}_I
    \end{autobreak}

    \\
    \begin{autobreak}
    \MoveEqLeft
    \mathcal{O}_{TR} = - 4\tilde{T}_{R} {\rm d}_t 
    + 2 \tilde{T}_R \tilde{T}_T
    + 2 \tilde{R}_P \tilde{R}_T
    \end{autobreak}

        \\
    \begin{autobreak}
    \MoveEqLeft
    \mathcal{O}_{RT} = 2 \tilde{T}_R \tilde{T}_T
    - 4 \tilde{R}_T + 2 \tilde{R}_{R} \tilde{R}_T
    \end{autobreak}
\end{align}
and $\mathcal{O}_{IR} = \mathcal{O}_{RI} = 0$. The full operator is constructed using \eqref{eqn:Multi-equation-operator-definition}.

\subsubsection*{The Sachs Equations}
For the Sachs equations, we construct $\mathcal{O}_{X_j X_k}$, where $X_j \in \{\theta, \omega, \sigma_+, \sigma_\times \}$. Then we have
\begin{align}
    \begin{autobreak}
    \MoveEqLeft
        \mathcal{O}_{\theta \theta} = 3\theta^{2} + 2\theta' - 4\omega^2 + 4\sigma^2 
        + 4\theta{\rm d}_{\lambda} - 2 {\rm d}_{\lambda}^{2} 
        + 2\mathcal{R}
        + 2\omega^2
        + 2\sigma_{+}^{2}
        + 2\sigma_{\times}^{2}
    \end{autobreak}

    \\
    \begin{autobreak}
    \MoveEqLeft
    \mathcal{O}_{\omega \omega} = -4(\theta^{2}+2\theta'-12\omega^2+4\sigma^2 + 2\mathcal{R})
    + 2(\theta^2+2\theta{\rm d}_{\lambda}-{\rm d}_{\lambda}^{2})
    \end{autobreak}

    \\
    \begin{autobreak}
    \MoveEqLeft
        \mathcal{O}_{\sigma_{+} \sigma_+} = 4(\theta^2+2\theta'-4\omega^2+12\sigma_{+}^2+4\sigma_{\times}^2+ 2\mathcal{R})
        + 2(\theta^2+2\theta{\rm d}_{\lambda}-{\rm d}_{\lambda}^{2})
    \end{autobreak}

    \\
    \begin{autobreak}
    \MoveEqLeft
        \mathcal{O}_{\sigma_{\times} \sigma_{\times}} = 4(\theta^2+2\theta'-4\omega^2+4\sigma_{+}^2+12\sigma_{\times}^{2}
        + 2\mathcal{R}) + 2(\theta^2+2\theta{\rm d}_{\lambda}-{\rm d}_{\lambda}^{2})
    \end{autobreak}

    \\
    \begin{autobreak}
    \MoveEqLeft
        \mathcal{O}_{\omega \theta} = -8\theta\omega-16\omega{\rm d}_{\lambda}
        + 2(2\theta\omega+\omega')
    \end{autobreak}

    \\
    \begin{autobreak}
    \MoveEqLeft
        \mathcal{O}_{\theta \omega} = -8\theta\omega 
        + 2(2\theta\omega+\omega'+2\omega{\rm d}_{\lambda})
    \end{autobreak}

    \\
    \begin{autobreak}
    \MoveEqLeft
        \mathcal{O}_{\sigma_+ \theta} = 8 \theta \sigma_+ + 16 \sigma_+ {\rm d}_\lambda 
        + 2(2\theta\sigma_{+}+\sigma_{+}'+\mathcal{C}_{+})
    \end{autobreak}

    \\
    \begin{autobreak}
    \MoveEqLeft
        \mathcal{O}_{\theta \sigma_+} = 8\theta\sigma_{+} + 2(2\theta\sigma_{+}
        + \sigma_{+}'+2\sigma_{+}{\rm d}_{\lambda}+\mathcal{C}_{+})
    \end{autobreak}

    \\
    \begin{autobreak}
    \MoveEqLeft
        \mathcal{O}_{\sigma_\times \theta} = 8 \theta \sigma_\times + 16 \sigma_+ {\rm d}_\lambda 
        + 2(2\theta\sigma_{\times}+\sigma_{\times}'+\mathcal{C}_{\times})
    \end{autobreak}

    \\
    \begin{autobreak}
    \MoveEqLeft
        \mathcal{O}_{\theta \sigma_\times} = 2(2\theta\sigma_{\times}+\sigma_{\times}'
        +2\sigma_{\times}{\rm d}_{\lambda}+\mathcal{C}_{\times})
    \end{autobreak}

    \\
    \begin{autobreak}
    \MoveEqLeft
        \mathcal{O}_{\omega \sigma_+} = -32\omega\sigma_{+}
    \end{autobreak}

    \\
    \begin{autobreak}
    \MoveEqLeft
        \mathcal{O}_{\sigma_+ \omega} = -32\omega\sigma_{+}
    \end{autobreak}

    \\
    \begin{autobreak}
    \MoveEqLeft
        \mathcal{O}_{\omega \sigma_\times} = -32\omega\sigma_{\times}
    \end{autobreak}

    \\
    \begin{autobreak}
    \MoveEqLeft
        \mathcal{O}_{\sigma_\times \omega} = -32\omega\sigma_{\times}
    \end{autobreak}

    \\
    \begin{autobreak}
    \MoveEqLeft
        \mathcal{O}_{\sigma_+ \sigma_\times} = 32\sigma_{+}\sigma_{\times} 
    \end{autobreak}

    \\
    \begin{autobreak}
    \MoveEqLeft
        \mathcal{O}_{\sigma_\times \sigma_+} = 32\sigma_{+}\sigma_{\times} 
    \end{autobreak}
\end{align}
The full operator is constructed using \eqref{eqn:Multi-equation-operator-definition}.

\section{Parameters used for simulation of the Amari model}
\label{app:amari}
In the main text, we simulated the neural field equations, based on the Amari framework model with parameters from \cite{taylor2014computational}. For the timescales, $t_1 = 26, t_2=1.25t_1, t_3 = 0.1 t_1, t_4=0.1t_1$. For the input strengths, we took $h_P = -0.35, h_I=-3.4, h_T=-2, h_R=-5$. For the connectivity strengths, we took $c_1 = 1.8, c_2=4, c_3 = 1.5, c_4=0.2, c_5 = 10.5, c_6=0.6, c_7=3, c_8=3, c_9=1$. We used activation function parameters $\epsilon=250000, a=2.8, b=0.5$.

\section{Transitions in the Sachs equations}
\label{app:sachs}
In the main text, we showed that the Sachs system \eqref{eq:JW-theta}-\eqref{eq:JW-sigma_times} can produce solutions which blow up in finite time provided that $\sigma_0^2 \geq \omega_0^2$, and either 
\begin{itemize}
    \item $\theta_0 \leq 0$, or
    \item $\theta_0 > 0$ and $\theta_0^2 < 4(\sigma_0^2 - \omega_0^2)$.
\end{itemize}
Below, we provide a derivation of these conditions.
Firstly, consider the special case in which both $\omega$ and $\sigma$
are initially zero. The transition between regular and singular evolution can be
predicted using the focusing theorem~\cite{Poisson_2004}. In this case, the system \eqref{eq:JW-theta}-\eqref{eq:JW-sigma_times}
reduces to:
\begin{equation}
    \dot \theta+\frac{1}{2}\theta^{2}(\lambda)=0,
\end{equation}
which is solved by
\begin{equation}
    \theta(\lambda)=\frac{2}{2\theta_{0}^{-1}+\lambda}.
\end{equation}
This solution has a finite-time singularity at $\lambda=2 \, |\theta_0|^{-1}$
if $\theta_0<0$, and is regular at all times if $\theta_0 \geq 0$.

In the general case, we notice that:
\begin{equation}
    \label{eq:JW-omega-sigma-relation}
    \frac{\omega}{\omega_0} = e^{-\int_0^\lambda \theta(\lambda'){\rm d} \lambda'} = \frac{\sigma}{\sigma_0},
\end{equation}
This implies that solution trajectories cannot cross the 
surface of the $\omega^2=\sigma^2$ surfaces, and that the term
$2 (\sigma^{2}(\lambda) -\omega^{2}(\lambda))$ in (\ref{eq:JW-theta})
cannot change sign during the evolution. On the surface
of this cone, $\omega$ and $\sigma$ will generally be different from 
zero, but they will not enter the equation for $\theta$, so the 
focusing theorem still applies and the light curves will be regular
for $\theta_0 \ge 0$ and singular for $\theta_0<0$, regardless of 
$\omega_0$ and $\sigma_0$.

Away from these special surfaces, if $\omega_0 \neq 0$,
equation (\ref{eq:JW-theta}) can be combined with \eqref{eq:JW-omega} and \eqref{eq:JW-omega-sigma-relation} to give:
\begin{equation}
    {\ddot \omega} = \frac{3}{2} \frac{\dot \omega^2}{\omega}+2\omega^3 \xi_\omega.
\end{equation}
Similarly, if $\sigma_0 \neq 0$, we have:
\begin{equation}
    {\ddot \sigma} = \frac{3}{2} \frac{\dot \sigma^2}{\sigma}+2\sigma^3 \xi_\sigma,
\end{equation}
with $\xi_\omega = \sigma_0^2/\omega_0^2-1=\xi_\sigma/(1-\xi_\sigma)$.
The solutions for ODEs of this form can be found analytically to be:
\begin{eqnarray}
    \label{eq:oml}
    \omega(\lambda) &=& \frac{\omega_0}{(\frac{\theta_0^2}{4}-\xi_\omega \omega_0^2)\lambda^2 + \theta_0 \lambda + 1}, \\
    \label{eq:sigl}
    \sigma(\lambda) &=& \frac{\sigma_0}{(\frac{\theta_0^2}{4}-\xi_\sigma \sigma_0^2)\lambda^2 + \theta_0 \lambda + 1}.
\end{eqnarray}

The existence of finite-$\lambda$ singularities can be shown
by examining the roots of the denominators in (\ref{eq:oml})-(\ref{eq:sigl}), given by:
\begin{equation}
    \lambda_{\pm} = \frac{2 \left(-\theta _0 \pm 2 \sqrt{(\sigma _0^2 - \omega _0^2)}\right)}{\theta _0^2- 4( \sigma_0^2 - \omega _0^2 )}
\end{equation}
We can therefore deduce that:
\begin{itemize}
    \item If $\theta_0 \leq 0$, $\lambda_{-}$ is always real and positive
when $\omega_0^2 \leq \sigma_0^2$, and no roots are real if $\omega_0^2 > \sigma_0^2$.
    \item If $\theta_0>0$, $\lambda_{-}$ is real positive if $4 (\sigma_0^2-
    \omega_0^2) > \theta_0^2$, and no other roots are real positive if $4 (\sigma_0^2-
    \omega_0^2) \leq \theta_0^2$,
\end{itemize}
which are the conditions stated in the main text.

\section{How does $\tro$ compare with other tools used to evaluate the behaviour of differential equations?}
\label{sec:tro-sensitivity-evaluation}
In the main text, we proposed $\tro$ as a score used to evaluate the qualitative content of ordinary and partial differential equations. Other methods, such as calculation of the parametric sensitivity and solution entropy, may also be appropriate as a measure of solution behaviour. The parametric sensitivity provides a measure of the change in solution behaviour by evaluating parametric derivatives between solution points, and can be evaluated per solution point, or integrated over the whole solution domain. The solution entropy measures the spread of the power spectrum of a given solution, integrated up to a maximum wavenumber in Fourier space \cite{gleiser_entropic_2012}.

To evaluate the relative power of $\tro$ with respect to such alternatives, we performed additional calculations for the nonlinear pendulum (\ref{eq:pendulum-ode}) and Allee reaction-diffusion equation (\ref{eqn:allee-spatial}). For parameter $\mu$, we define the global parametric sensitivity $\Gamma(\mu)$
between two solutions $u(x^b; \mu)$ as
\begin{equation}
    \Gamma[u(x^b, \mu)] = \int\limits_{\Omega \times [t_i, t_f]} \left | u(x^b; \mu + \mathrm{\Delta} \mu) - u(x^b; \mu)\right |^2 {\rm d}x^b,
\end{equation}
where $x^b$ represent the appropriate spacetime coordinates. For the solution entropy, following~\cite{gleiser_entropic_2012} we define the power spectrum of a solution as
\begin{equation*}
    \hat{U}(k; \mu) = \left | \int_{\Omega \times [t_i, t_f]} u(x^b; \mu) e^{- i k^b x^b} {\rm d}x^b \right |^2,
\end{equation*}
the modal fraction as
\begin{equation*}
    \hat{u}(k; \mu) = \frac{\left | \hat{U}(k ; \mu)\right |^2}{\int_{K}{ \left | \hat{U}(k; \mu)\right |^2 {\rm d}k^b}},
\end{equation*}
and the solution entropy as
\begin{equation}
    S_c[\hat{u}(k; \mu)] = - \int_{K} \hat{u}(k; \mu) \ln(\hat{u}(k; \mu)) {\rm d}k^b,
\end{equation}
where $K$ denotes the region of integration in frequency space associated with $\Omega \times [t_i, t_f]$. In calculation of the parametric sensitivity, for the pendulum we took $\Delta \dot{\theta_0} = 0.002$, and for the Allee RDE we took $\Delta \alpha = 0.02$.

\begin{figure}[!b]
    \centering
    \includegraphics[width=0.45\textwidth]{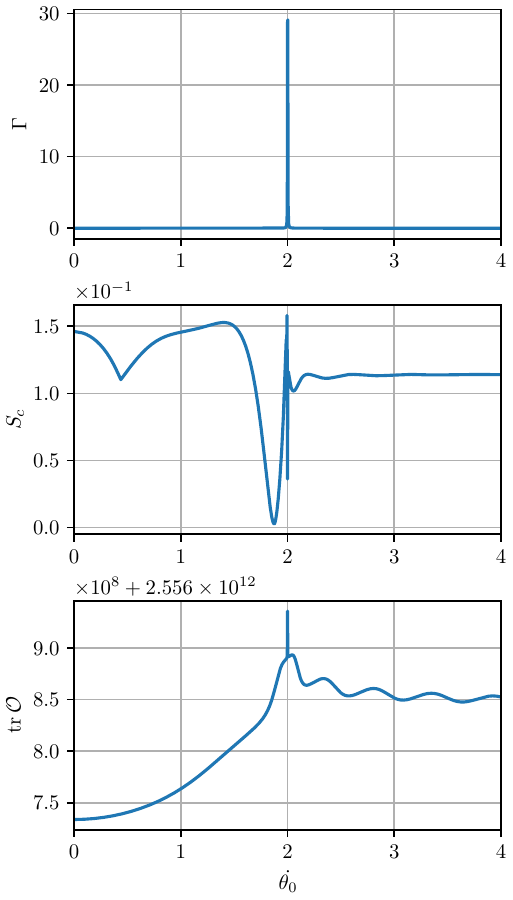}
    \caption{Comparison between $\tro$, the solution entropy $S_c$, and parametric sensitivity $\Gamma$ for the pendulum as a function of $\dot{\theta_0}$.}
    \label{fig:sensitivity-comparisons-pendulum}
\end{figure}

The results for the pendulum are shown in Figure \ref{fig:sensitivity-comparisons-pendulum}. We found that each metric highlights different features of the solution space; the parametric sensitivity is able to detect the separatrix solution, but does not change appreciably between the different regimes detected. By comparison, the solution entropy can detect the special solution, but exhibits additional peaks between the $\dot{\theta_0}=0$ and the special solution at $\dot{\theta_0} = 2$ which are not attributable to a qualitative change in solution behaviour. $\tro$ is able to successfully detect the special solution, and takes on qualitatively different values on either side of the transition. However, it also exhibits additional, smaller peaks in the $\dot{\theta_0} > 2$ regime.

\begin{figure}[!t]
    \centering
    \includegraphics[width=0.45\textwidth]{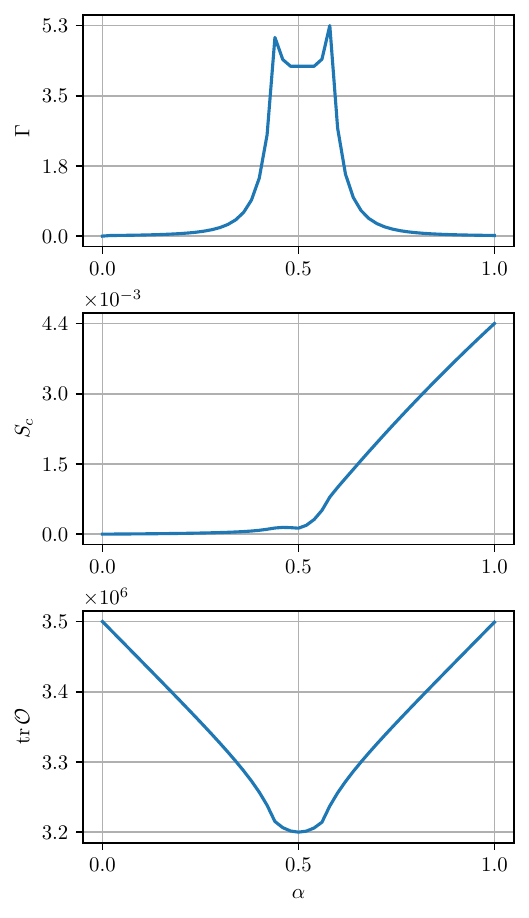}
    \caption{Comparison between $\tro$, the solution entropy $S_c$, and parametric sensitivity $\Gamma$ for the Allee reaction-diffusion equation as a function of $\alpha$.}
    \label{fig:sensitivity-comparisons-allee}
\end{figure}

The results for the Allee reaction-diffusion equation are shown in Figure \ref{fig:sensitivity-comparisons-allee}. Similarly to the pendulum, each metric exhibited different properties. The parametric sensitivity rises as it approaches the special solution $\alpha = 0.5$ from either direction, which is represented by a minima. The parametric sensitivity shows additional structure around $\alpha = 0.44$ and $\alpha = 0.56$. These solutions are the furthest in parameter space from the special solution at $\alpha = 0.5$ whilst still not achieving complete migration ($u=1$ everywhere on the domain) or extinction ($u=0$ everywhere on the domain) by the end of the simulation time. Conversely, the entropy does contain a minimum at the special solution, but the minima is shallow with respect to the overall landscape, which decreases to a global minimum at $\alpha = 0$. Comparatively, $\tro$ has a clear minima at the special solution, and shows a change in curvature around the solutions with partial migration or extinction.

From these observations, we conclude that all three measures are of use in detecting qualitative changes in solution character. We note that the solution entropy is the least interpretable, given that the transitions we observe are not expected to radically change the power spectrum of the solutions. By comparison, $\tro$ and the parametric sensitivity contain some overlapping information, with each reliably detecting the special solutions. Depending on the application, $\tro$ may provide additional information due to its different scales over different regimes. Additionally, it possesses some long range character which could be used to more efficiently search parameter space, although to be employed most effectively, further work will be needed to understand the finer structure of its behaviour. 

\vfill

\bibliography{biblio}

\end{document}